# Enhancing transparency in AI-powered customer engagement

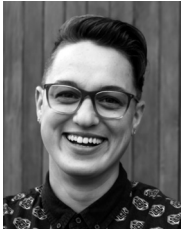

**Tara DeZao**
Director of Product Marketing, AdTech and MarTech, Pegasystems, USA

Tara DeZao, Director of Product Marketing, AdTech and MarTech at Pegasystems, has had a successful career in the marketing departments of both start-ups and Fortune 500 enterprise technology companies including Oracle and EMC. As a subject matter expert in the field of advertising and marketing, she is an experienced thought leader across written, audio and live mediums. She holds a BA from the University of California at Berkeley and an MBA from the University of Massachusetts at Amherst.

E-mail: tara.dezao@pega.com

**Abstract**   This paper addresses the critical challenge of building consumer trust in AI-powered customer engagement by emphasising the necessity for transparency and accountability. Despite the potential of AI to revolutionise business operations and enhance customer experiences, widespread concerns about misinformation and the opacity of AI decision-making processes hinder trust. Surveys highlight a significant lack of awareness among consumers regarding their interactions with AI, alongside apprehensions about bias and fairness in AI algorithms. The paper advocates for the development of explainable AI models that are transparent and understandable to both consumers and organisational leaders, thereby mitigating potential biases and ensuring ethical use. It underscores the importance of organisational commitment to transparency practices beyond mere regulatory compliance, including fostering a culture of accountability, prioritising clear data policies and maintaining active engagement with stakeholders. By adopting a holistic approach to transparency and explainability, businesses can cultivate trust in AI technologies, bridging the gap between technological innovation and consumer acceptance, and paving the way for more ethical and effective AI-powered customer engagements.

## INTRODUCTION

Businesses are leveraging artificial intelligence (AI) to gain scale, increase operational efficiency and improve customer and user experiences. It is becoming a competitive differentiator and will soon become a minimum requirement to vie for market share. Despite AI's capacity to automate administrative tasks, remove friction from the customer journey and elevate employee skills and experiences, both consumers and organisational leaders remain wary of utilising it. According to a survey of 2,000 employed Americans by market research company OnePoll, commissioned by *Forbes Advisor*, over 75 per cent of those surveyed were concerned about misinformation driven by AI in their interactions with businesses in content such as websites and product information.[1]

While some users distrust AI, many consumers do not even know how frequently they use it. A 2023 study by Pew Research found that 44 per cent of US consumers surveyed do not think they are encountering AI on a regular basis.[2] In 2024 Pew reports that at least 90 per cent of US consumers own a smartphone.[3] Both Android and Apple smartphones are enabled with voice assistants out-of-the-box, which suggests a high number of consumers are not aware of their frequent AI use.

Although businesses continue to expand the integration of AI into all facets of commerce and society, even those overseeing it cannot always understand why AI systems make certain decisions. And if subject matter experts cannot, businesses cannot expect consumers to understand how it works and therefore trust it. Trusting technology is essential to adopting and accepting it. The convergence of consumers, business and AI will require business to prioritise transparency and accountability with these powerful technologies.

## EXPLAINING AI-POWERED DECISION MAKING IN CUSTOMER ENGAGEMENT

Fundamentally, an algorithm is a set of instructions that an AI system is expected to follow to perform a task. The model is trained using various kinds of data learn like a human brain. A basic example would be an AI-powered chatbot that is trained with volumes of data to field and deflect basic customer questions on an e-commerce website. If a customer asks a question to the chat such as 'How can I cancel my subscription?', the chatbot has been taught where to direct that customer by training it with sample interactions and data over and over.

This is a basic example used every day by brands and distributors that operate call centres to reduce the cost of an interaction where human labour is not needed to solve a simple problem. That said, we also ask AI algorithms to make more complex decisions, and as the complexity increases the decision-making process can become less explainable to a human audience — more opaque. AI explainability, also referred to as 'transparency', is a critical pillar that that defines responsible AI practices. Transparency is one component that can inspire trust in both consumers and organisational leaders. It prevents us from running afoul of social norms and helps to avoid ethical bias that can inadvertently creep into algorithms.

Transparency is an important trait within technology that ensures fairness in AI-powered decisioning for those who are affected by those decisions, such as consumers. For example, predictive AI models that make future predictions of a consumer's behaviour based on historical data are used every day by apps, employers, advertisers, lenders, insurers and beyond.[4] It is important to be able to understand why, for example, a lender offers a credit card to one consumer but not another when they have similar demographic information and credit scores (see Figure 1).

The explanation may be based on legitimate determinants. In the above example Jason might not have received an offer because he already had that product. But he also might not have received that offer because of bias in the data fields that are used to determine a consumer's eligibility to receive an offer from a lender, such as age, race, gender, region, etc.

The only way to determine this would be to review the decision criteria the algorithm is based on, and to do that, the algorithm would need to be transparent. According to Kearns and Roth in *The Ethical Algorithm*,

> A hammer might be put to an unethical use … Anything ethical about the misuse of a hammer can be attributed to the human being that wields it. But algorithms — especially via machine learning are different … because we allow them a significant amount of agency to make decisions without human intervention and because they are often so complex and opaque, that even their designers cannot anticipate how they will behave in many situations.[5]

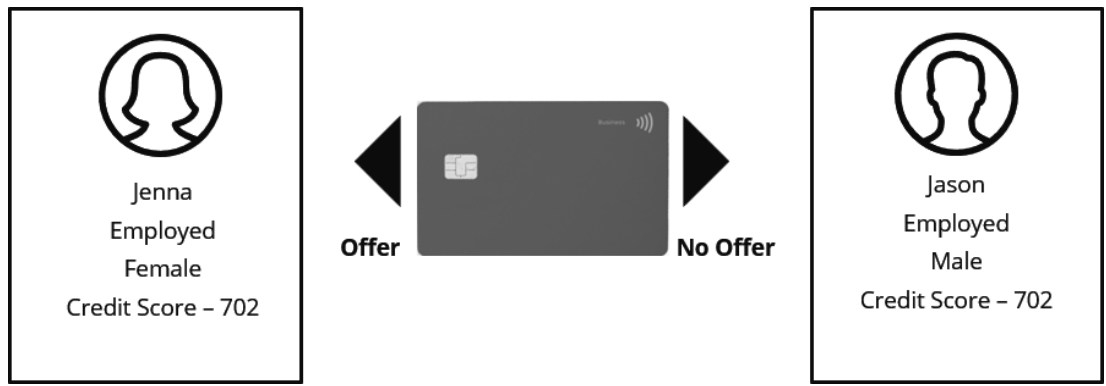

**Figure 1:** Credit score eligibility

The transparency required for AI applications can differ significantly across sectors, particularly in tightly controlled fields where adherence to regulation is critical. Additionally, innovation often outpaces regulation and risks, while rules and enforcement vary widely based on things like geographic location.

Ultimately, building consumer trust through transparency cannot be forced by regulation, although it does help. Organisations will need to self-regulate by developing strategies and policies that demonstrate a commitment to transparency, fairness and accountability. Within the marketing and customer engagement use case, while opacity might be a requirement in industries like financial services, you would want to apply high levels of transparency to any decision that has the possibility for bias to enter the decision-making process. Lower levels of transparency might be more appropriate for an AI-powered decision that is responsible for generating and delivering creative artefacts such as imagery for advertising units. Both levels of transparency can exist within the same organisation for different programs.

Not following these practices comes with both reputational and financial dangers for brands. Apple and their finance partner Goldman Sachs were accused of gender bias by applicants for the then newly released Apple Card in 2019. There was a discrepancy in the amount of credit adults in the same household were offered and without algorithmic transparency, the brands were unable to explain the decision-making process initially. Ultimately they were investigated by the New York State Department of Financial Services (NYSDFS) and cleared of wrong-doing. NYSDFS investigators conducted a statistical analysis of almost 400,000 New York applicants, which showed that the models and algorithms Goldman Sachs uses to filter applicants 'did not consider prohibited characteristics of applicants and would not produce disparate impacts';[6] however, this took approximately two years to investigate and clear. While they ultimately broke no laws, they did fracture consumer trust and create poor customer experiences. Goldman Sachs announced at the end of 2023 that they were looking for solutions to exit the partnership because it was too risky and not profitable.[7]

## CHALLENGES TO ACHIEVING TRANSPARENCY

Developing more explainable AI models is the core tactic for achieving transparency, but that is often easier said than done. And many view AI models and algorithms as a secret sauce that, if exposed, would be tantamount to ceding competitive advantage. Algorithms are classified by some as intellectual property.

There is also a relationship between opacity and predictive power. More opaque models are often more powerful. As a marketer, I look at this as a similar comparison to the relationship between audience reach and accuracy in data-driven campaigns. The wider my audience is, the less relevant my messaging might be, whereas if my audience is more granular, I reach fewer people but the messaging may resonate more. Think podcast advertisement versus short message service (SMS) text message. It is a trade-off we must analyse against our goals and budgets.

When it comes to statistical and machine learning (ML) models, they range from simple and transparent to complex and opaque. Some AI models are incredibly complex and difficult to interpret. Two examples of more complex models are deep neural networks (DDN) and Bayesian networks. Some examples of technology that uses DNNs are voice assistants like Siri and Alexa, recommendation algorithms like those used by Netflix and YouTube, language translation services and self-driving cars.

Examples of more simplistic models are linear regression and decision trees (DTs). A DT can be made on a simple piece of paper by someone who is not a data scientist. I could make one right now to figure out where I am going to go on vacation this year. It is extremely easy to see the decision path and how you arrive at an outcome. DTs can be used for medical diagnosis and loan approval processes; linear regression is used in credit scoring and real estate pricing.

You will notice the trade-off between accuracy and opacity. My Netflix recommendations are going to be a lot more accurate, more of the time, than a human using a DT to figure out if I have a virus. And although there is an algorithm that is widely used for real estate appraisals, the process is variant based on factors outside of the model, including who is performing the evaluation. The relationship between data, algorithms and human intervention is a complex one, and there are ways to make them work together more effectively to increase both transparency and accuracy.

## GO BEYOND ALGORITHMS TO ENHANCE TRANSPARENCY

In addition to algorithmic transparency, there are many ways to create a holistic transparency strategy across the enterprise to increase both employee and customer trust of AI systems. Combining cultural tenants with technological transparency will help achieve those trust outcomes.

### Create a culture of organisational accountability

Paramount in combatting challenges to achieving transparency is for organisations to create a culture of accountability. Accountability should be shared, and not just taken on by technologists but from all functional areas, for example marketing, operations, sales, customer service and beyond. Making this value part of the culture reinforces its importance to business.

This requires investment and commitment by organisational leadership. According to Greg Khilstrom, author of *The Agile Brand Guide to Generative AI*, 'To address these challenges effectively, companies must be willing to invest in proper training and education for their employees, ensure the quality of the data used in the AI systems and stay up-to-date with the latest legal and regulatory developments'.[8]

### Keep employees educated and up to date with the rapidly evolving AI landscape

The field is continuously changing, sometimes by the minute; however, it is important to keep the workforce informed about the developments in this area (see Figure 2), especially as related to policies and regulations that will affect both the organisation and the customer. Pay attention to the emerging trends.

**Be open and transparent about data policies and processes**

Data is the lifeblood of an AI system. While it is required by law in many locations, be transparent about how the data your organisation holds is collected, processed and used within the organisation and post it so that consumers and employees can access this information easily *and* understand it. According to a recent study by the International Association of Privacy Professionals, the chief step that companies can take to assuage consumer privacy concerns is to provide clear and understandable privacy information on how their data is being processed.[9]

An example of a brand that publicises its privacy policies and responsible AI pillars in a way that is both accessible and easy for consumers, employees and stakeholders to understand is Intel. Its data and privacy policies are actually embedded in an overall framework published on its website in the section 'Responsible AI Principles':

> Intel supports privacy rights by designing our technology with those rights in mind, including being transparent about the need for any personal data collection, allowing user choice and control; and, to design, develop and deploy our products with appropriate guardrails to protect personal data.[10]

This is one of seven principles the company has laid out as paramount for the design, development and responsible use of AI. Within these seven principles it also features commitments to 'enable transparency and explainability' as well as 'enable human oversight'.

*Human oversight*

It is notable that in the above example, Intel mentions human oversight as a pillar of responsible technology use because it is a critical tenant in maintaining AI transparency. And as mentioned above, often the high opacity/high efficacy trade-off of AI algorithms means that they can be so complex that humans cannot understand them and thus they are allowed to operate with minimal human oversight, which is a feature of their design to intentionally obscure their inner workings from straightforward interpretation.

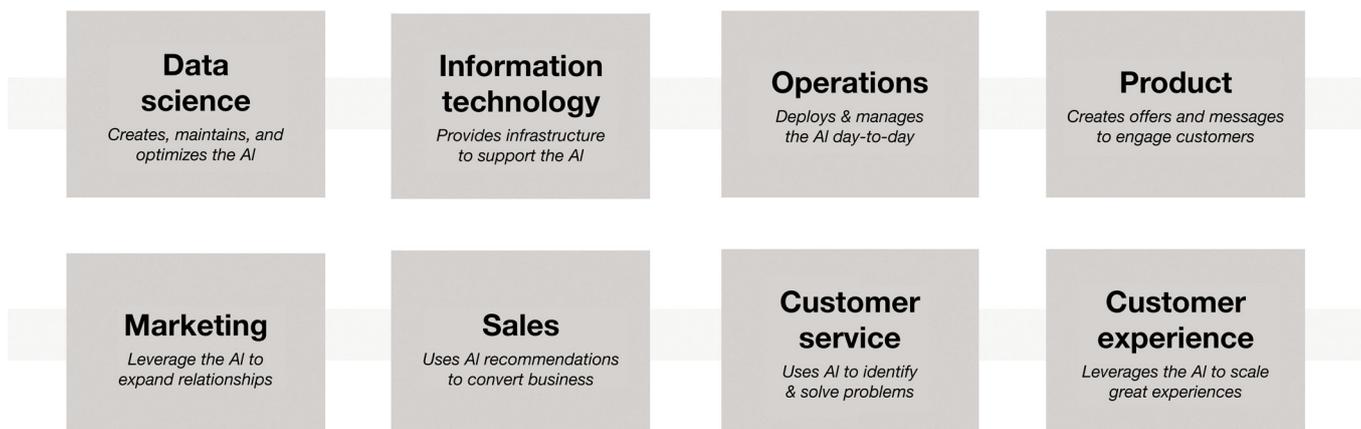

**Figure 2:** Who has accountability for responsible use of AI?

A human auditor to oversee decisions and performance can instil a high level of accountability for AI-powered decisions. If there is a problem or bias emerges, a human auditor can catch it and intervene early before it is reinforced over and over. There are two pathways: human-in-the-loop and human-over-the-loop. With 'human-in-the-loop' the AI cannot decide without it being checked by a human. This would be ideal for high-stakes situations such as medical diagnoses. A 'human-over-the-loop' scenario is one where if a human disagrees with an AI-powered decision, they can override it. This would be appropriate for AI that is lower risk, an example being a traffic prediction system looking for the shortest route to a destination.[11] It is easy for a human to override this decision.

Additionally, we are starting to see the intellectual property defence of opacity crumble in the face of AI malfunctions that cause public embarrassment for brands. Consider, for example, when Google released their AI chatbot Bard in a public demo in early 2023. When questioned, Bard returned factually incorrect information, causing Google's parent company Alphabet to lose US$100bn in market value in one day when it could not or would not explain why. Meta, parent company of Facebook, Instagram and WhatsApp, has been notoriously opaque around its models, defending as intellectual property. After Google's mishap, however, later that year Facebook launched its AI model, LLaMA 2, and will allow the wider AI community model to download it and make changes.

In an *MIT Technology Review* article from July 2023, Melissa Heikkila writes, 'This could help make it safer and more efficient. And crucially, it could demonstrate the benefits of transparency over secrecy when it comes to the inner workings of AI models'. She adds:

> If products built on much-hyped proprietary models suddenly break in embarrassing ways, and developers are kept in the dark as to why this might be, an open and transparent AI model with similar performance will suddenly seem like a much more appealing—and reliable—choice. Meta isn't doing this for charity. It has a lot to gain from letting others probe its models for flaws. Ahmad Al-Dahle, a vice president at Meta who is leading its generative AI work, told me the company will take what it learns from the wider external community and use it to keep making its models better.[12]

This underscores that what is ethically ideal and beneficial for developers and users is also good for business.

*Test and learn*

Related to human oversight is testing and learning. Simulating various scenarios with different products, conversations, customer profiles, offers and beyond before putting them into production in programs that touch customers can help us understand how a program or algorithm would work in real life, but within a test environment. AI and simulation technologies have each gained and sustained momentum on their own for almost ten years. Recently, engineers are recognising significant benefits at the convergence of these fields, due to the complementary nature of their advantages and disadvantages.[13]

Brands can use data from customer profiles modelled after their existing customer base, samples from of production data and customer interaction history to test performance in a separate cloud environment or sandbox. Then, data scientists review the outcomes and tweak them if they find a recurring pattern that could trigger questions about transparency, before using algorithms to drive interactions with real human customers.

**Do not rely on legislation — self-regulate**

Risk mitigation with AI systems is not just an essential strategy to prevent damage to brand equity; in some locales and industries, it is the law. Legislating these systems without constraining innovation is a net positive for society. What is challenging, however, is that there is a wide-ranging appetite to pass uniform rules and standards.

Take, for example, the Artificial Intelligence Act in the European Union (EU), which is legislation aimed at regulating the development, deployment and use of AI within the EU. The main goal of the Act is to ensure that AI systems are safe, transparent and accountable, and that they respect existing laws on privacy and fundamental rights.[14] It is the first law of its kind. But passing something like this in the US would be nearly impossible given that the legislative environment is not as friendly to meaningful technology regulation. In the absence of federal law, all 50 states would be required to pass their own laws, leading to a patchwork of varying rules and regulations, confusing and costly to enforce.

The United States Executive Order on Artificial Intelligence (AI) was issued by the White House in February 2019. The primary aim of this executive order is to promote and protect national AI technology and innovation, with a focus on maintaining the US global leadership position in AI research, development and deployment.[15]

While this executive order reflects the US government's recognition of the strategic importance of AI in economic growth, national security and technological advancement, it is not law or even meaningful regulation — it is an executive order that governs only the federal government's use of and approach to AI systems. Innovation and the evolution of technology will outpace even the most meaningful regulation. It is important for organisations to prioritise ethics and follow regulatory principals even when not mandated by law.

The journey towards embedding transparency in AI-powered customer engagement is both critical and complex. The value of AI in enhancing business operations and customer experiences is undeniable and its integration into society inevitable. The full potential of these technologies can only be realised, however, when there is a foundational level of trust between consumers and businesses. Achieving this trust requires a commitment to transparency, accountability and fairness in AI applications and the strategies that govern them. Organisations must navigate the delicate balance between leveraging the power of AI for competitive advantage and ensuring ethical, transparent practices that build consumer confidence and comply with evolving regulations. By prioritising explainable AI models, fostering an organisational culture of openness, and continuously engaging with stakeholders about the role of AI in their experiences, businesses can pave the way for a future where AI-powered customer engagement is not only effective but also trusted.

© Tara DeZao, 2024